# Nanoscale Chemical Heterogeneity Dominates the Optoelectronic Response over Local Electronic Disorder and Strain in Alloyed Perovskite Solar Cells

Authors

Kyle Frohna[1], Miguel Anaya[*,1], Stuart Macpherson[1], Jooyoung Sung[1], Tiarnan A. S. Doherty[1], Yu-Hsien Chiang[1], Andrew J. Winchester[2], Keshav M. Dani[2], Akshay Rao[1], Samuel D. Stranks[*,1,3]

[1] Cavendish Laboratory, University of Cambridge, Cambridge, UK

[2] Femtosecond Spectroscopy Unit, Okinawa Institute of Science and Technology Graduate University, Onna-son, Japan

[3] Department of Chemical Engineering and Biotechnology, University of Cambridge, Cambridge, UK

*sds65@cam.ac.uk, ma811@cam.ac.uk

**Halide perovskites perform remarkably in optoelectronic devices including tandem photovoltaics. However, this exceptional performance is striking given that perovskites exhibit deep charge carrier traps and spatial compositional and structural heterogeneity, all of which should be detrimental to performance. Here, we resolve this long-standing paradox by providing a global visualisation of the nanoscale chemical, structural and optoelectronic landscape in halide perovskite devices, made possible through the development of a new suite of correlative, multimodal microscopy measurements combining quantitative optical spectroscopic techniques and synchrotron nanoprobe measurements. We show that compositional disorder dominates the optoelectronic response, while nanoscale strain variations even of large magnitude (~1 %) have only a weak influence. Nanoscale compositional gradients drive carrier funneling onto local regions associated with low electronic disorder, drawing carrier recombination away from trap clusters associated with electronic disorder and leading to high local photoluminescence quantum efficiency. These measurements reveal a global picture of the competitive nanoscale landscape, which endows enhanced defect tolerance in devices through spatial chemical disorder that outcompetes both electronic and structural disorder.**

Metal-halide perovskites have shown tremendous success in solar cell and light emitting applications[1,2]. Single junction and tandem solar cells based on halide perovskites have reached above 25% and 29% power conversion efficiency, respectively[3-5]. Most state-of-the-art perovskite solar cells incorporate mixed composition absorber layers alloyed at the A and/or X-site of the $ABX_3$ perovskite structure[5-10]. This is particularly the case in the top cells for perovskite-silicon and all-perovskite tandems in which compositional engineering including bromide-iodide alloying at the X site is required to reach optimal wider bandgaps to maximize efficiency [11,12]. The highest performing and most reproducible compositions across the range of device applications are comprised of mixed compositions such as $FA_{0.79}MA_{0.16}Cs_{0.05}Pb(I_{0.83}Br_{0.17})_3$ (FA=formamidinium, MA=methylammonium) found through empirical device optimization [1,4,8,9,13-16]. While the specific cation and halide composition fractions may vary, the presence of mixed halide compositions is ubiquitous, particularly in tandem applications. Yet, the unprecedented performance of both single junction and tandem devices comes in spite of the fact that these halide perovskite layers are optoelectronically, chemically and structurally heterogeneous across multiple length scales [17-22].

It has been shown that the bulk stoichiometry of halide perovskite materials governs their optoelectronic performance and stability at the macroscale [8,9,13,23]. Recently, we used microscopy-based techniques to show that trap clusters on the length scale of tens of nanometers appearing at specific grain boundaries limit the performance of halide perovskite films by promoting non-radiative recombination [22,24,25], which limits open-circuit voltage in solar cells and light output in LEDs. Using scanning electron diffraction correlated with photoelectron emission microscopy, we showed that these nanoscale trap clusters are specifically present at grain boundaries between pristine halide perovskite grains and structurally inhomogeneous phases [22]. In other work the homogenization of nanoscale stoichiometry has been linked to beneficial device performance improvements[26]. In mixed halide perovskite systems with bromide concentrations greater than ~20%, halide segregation has been reported to occur under light and/or bias with the resulting iodide-rich, low-bandgap regions widely thought to act as non-radiative recombination sites and reduce device performance[27-29]. This is highlighted by a perceived voltage plateau in devices above this threshold bromide content[30]. In addition, structural variation in the form of strain has been linked to non-radiative recombination[31,32]. These results suggest that structural and/or chemical heterogeneity is detrimental to device performance, in line with classical understanding of semiconductor behavior. However, recent results have shown that the voltage loss in wide bandgap mixed-halide perovkite solar cells is limited by intrinsic material quality and that high voltages can be achieved in spite of ionic segregation[33]. Furthermore, macroscopic studies have shown that carrier accumulation of one carrier type in these mixed-cation, mixed-anion films due to photodoping of holes in lower bandgap regions promotes radiative charge carrier recombination[34], which would be beneficial to device performance. A global understanding of the influence of the nanoscale landscape on device performance, and how the apparently detrimental and beneficial disorder can be reconciled, will be crucial for progessing the field. We solution-processed mixed-cation, mixed-halide $FA_{0.79}MA_{0.16}Cs_{0.05}Pb(I_{0.83}Br_{0.17})_3$ perovskite films on glass, as well as solar cells in which the perovskite is sandwiched between contacts comprised of poly[bis(4-phenyl)(2,4,6-trimethylphenyl)amine (PTAA) and poly(9,9-bis(3'-(N,N-dimethyl)-N-ethylammoinium-propyl-2,7-fluorene)-alt-2,7-(9,9-dioctylfluorene))dibromide (PFN-Br) on the bottom side and phenyl-C61-butyric acid methyl ester (PCBM) on the top side [21], with the complete stack as glass/indium tin oxide (ITO)/PTAA:PFN-Br/perovskite/PCBM/bathocuproine (BCP)/Ag (see Supplementary Fig. 1). We developed a novel technique based on a wide-field, hyperspectral microscope to assess the important optoelectronic properties of the samples, namely absorption and luminescence, in a quantitative manner at the nanoscale. The setup is calibrated to measure the spectrally resolved, absolute number of photons leaving the sample locally at each point of the region of interest (see Figure 1a, and Methods). We assess local non-radiative recombination processes of bare perovskite films (see Supplementary Fig. 1 for equivalent measurements of a full device stack), which negatively impact the performance of solar cell devices [35], through extraction of absolute photoluminescence (PL) maps as a function of emission wavelength (cf. Figure 1b). The 405-nm continuous wave laser excitation intensity is set to ~90 mW/cm$^2$, equivalent to illumination with ~1.4 sun for this bandgap (see Methods), which emulates the conditions of a functioning solar cell device; we do not see considerable transient light soaking effects under these conditions (see Supplementary Fig. 2) [36,37]. The samples show PL spectra centred at approximately 1.62 eV (765 nm), with local variations in intensity, peak position and shape on the hundreds of nanometres scale (see example spectra in Figure 1b, taken from map in Figure 1d). By fitting the absolute PL spectra using the generalized Planck law [38], we extract the Quasi-Fermi Level Splitting (QFLS) from each point and plot this in Figure 1f (see Methods and Supplementary Fig. 3), which provides information about the excited state carrier distributions and gives an upper bound for the open circuit voltage of a solar cell for the device under analysis. In films, we find variations above 20 meV in the local QFLS values, including local performance hot spots with values above 1.24 eV, corresponding to ~93% of the radiative limit of ~1.32 eV for this bandgap (1.62 eV). This spatial heterogeneity is particularly evident in the full device stack (Supplementary Fig. 1) where QFLS values range from below 1.07 eV to above 1.11 eV, with the increased spread likely due to additional spatially varying carrier quenching from the contacts. The spatially averaged QFLS in the bare film corresponds to an optically implied open-circuit voltage of ~1.23 V, with the full device exhibiting a value of 1.08 V, consistent with non-radiative losses associated with the contacts [39,40] and comparable to the electrically measured value of 1.11 V.

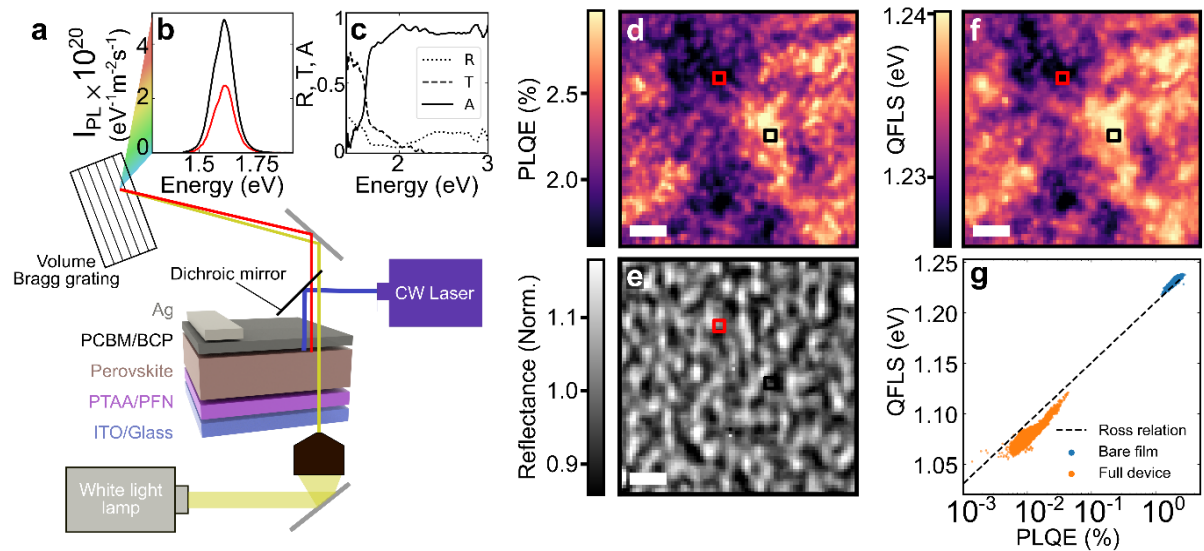

**Figure 1: Hyperspectral microscopy of perovskite solar cell device stacks.** a) Schematic of the hyperspectral microscopy setup used to characterize the $FA_{0.79}MA_{0.16}Cs_{0.05}Pb(I_{0.83}Br_{0.17})_3$ perovskite films and devices. b) PL spectra from the black and red highlighted regions in panels d, e and f. c) Reflectance (R), transmittance (T) and absorptance (A) spectra from the region highlighted in black in those panels. d) Absolute PLQE, e) broadband reflectance and f) QFLS maps of a $FA_{0.79}MA_{0.16}Cs_{0.05}Pb(I_{0.83}Br_{0.17})_3$ perovskite device stack (without the back metal contact). Reflectance is normalised to the mean value of the map. g) Scatter plot of QFLS versus PLQE extracted from the maps in panels (d) and (f) compared to the Ross relation. All scale bars represent 2 μm.

Additionally, we employed calibrated white light sources to extract local reflectance and transmittance spectra, and hence compute absorptance at each point, from the same region (Figure 1c, 1e and Supplementary Fig. 4). The macroscopic short-circuit current of full solar cells (~20 mA/cm$^2$, Supplementary Fig. 1) matches the maximum current one can expect from the spatially averaged local reflectance maps (Figure 1e) assuming that all the light that is not reflected is absorbed (Supplementary Fig. 1). The absorptance spectra provide the fraction of photons absorbed by the device at the 405-nm wavelength employed to obtain PL maps, allowing us to construct photoluminescence quantum efficiency (PLQE) maps (Figure 1d); to the best of our knowledge, this is the first report of local PLQE in photovoltaic absorbers achieved under solar operating conditions[41]. The spatially averaged value of the local PLQE of 2.2 % is in good agreement with standard macroscale methodologies (3.2 %) [42] (see Methods), validating our approach. We note that both values are from external measurements, dictated by specific film outcoupling properties [43], and there is small systematic underestimation of the PLQE from the local measurements due to light waveguided within the film or substrate or light emitted in the backward hemisphere that is not collected. The PLQE ($\eta$) is logarithmically related to the QFLS ($\Delta\mu$) of a material through the Ross relation [44],

$$\Delta\mu = \Delta\mu_{rad} + kT\ln(\eta)$$

where $\Delta\mu_{rad}$ is the radiative limit QFLS for a given bandgap, $k$ is the Boltzmann constant and $T$ is temperature. We compare the local QFLS and PLQE extracted from each point as shown in Figure 1g and find that their relationship matches well with the Ross relation. Therefore, our quantitative optical microscopy represents a self-consistent picture of local optoelectronic properties that links directly to device performance metrics.

We now seek to understand how the heterogeneous sample properties impact other local optoelectronic parameters. Figure 2a and 2b shows PLQE and Urbach energy ($E_U$) maps, respectively, extracted from the same area of a thin film. In order to probe electronic disorder we extract the local $E_U$ from the red tail of the PL spectrum at each pixel [45] (see Methods and Supplementary Fig. 5); a low $E_U$ is associated with clean semiconducting behavior[45-47]. As shown in the 2-dimensional kernel density estimation plot in Figure 2c, there is a clear spatial anti-correlation between the PLQE and the $E_U$ (Spearman's r-value = -0.44, p-value << 0.0001), highlighting the regions that are the most emissive also show the cleanest sub-bandgap tail, consistent with what we would expect for a semiconductor. Figure 2d shows the spatially averaged PL spectra from the highest PLQE (>80th percentile) and lowest PLQE (<20th percentile) regions. These spectra show that the highest PLQE regions exhibit a distinctly red-shifted (low-energy) shoulder in the emission, which is absent in the low PLQE regions. By contrast, there is negligible difference in spectra between low and high PLQE regions in analogous single halide $FA_{0.79}MA_{0.16}Cs_{0.05}PbI_3$ thin film controls, and luminescence yields are an order of magnitude lower than the mixed halide counterparts (Supplementary Fig. 6). These quantitative hyperspectral microscopy results thus suggest that the mixed halide composition and its effects on the local energetic landscape and structure plays a key role in the observed optoelectronic quality of the perovskite films.

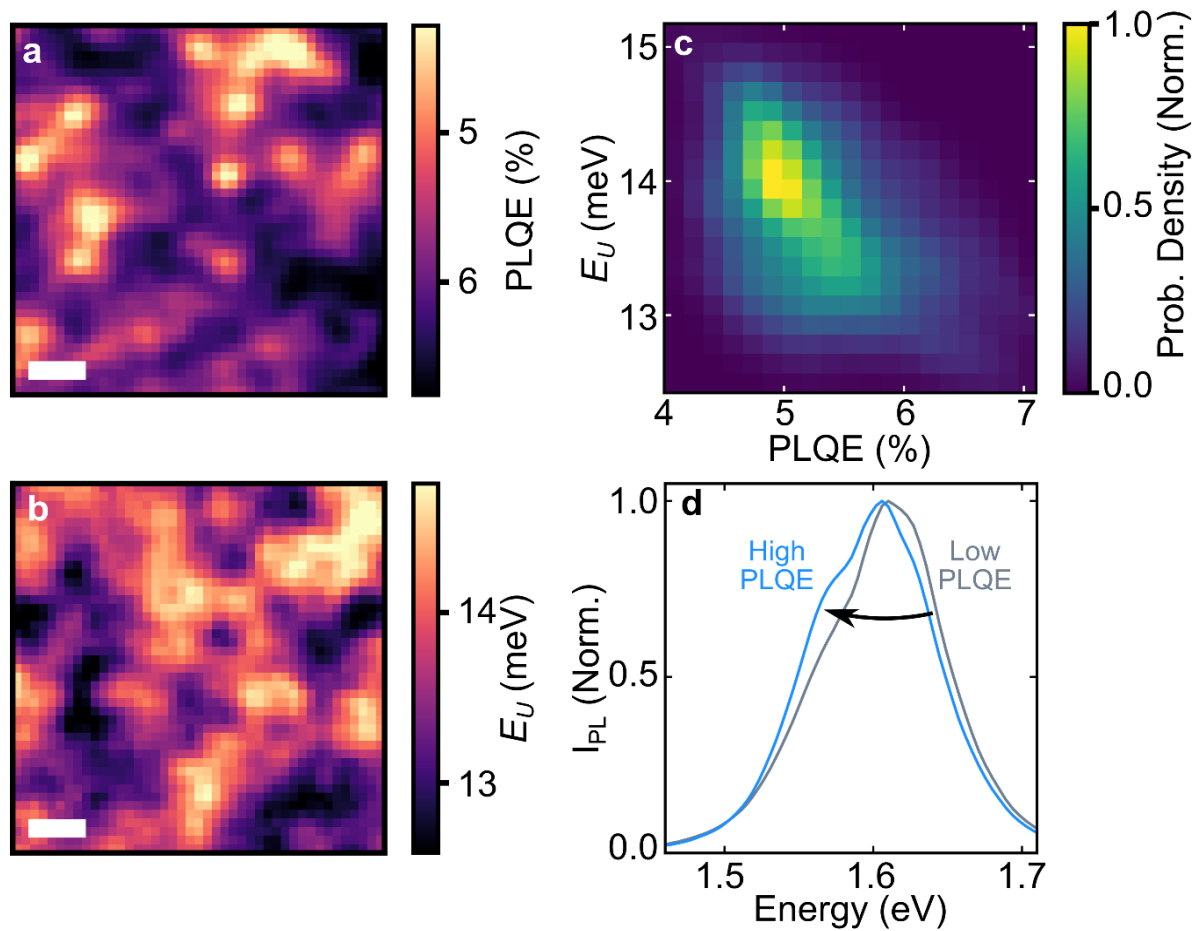

**Figure 2: Correlation between optoelectronic properties in FA$_{0.79}$MA$_{0.16}$Cs$_{0.05}$Pb(I$_{0.83}$Br$_{0.17}$)$_3$ perovskite films.** a) PLQE map and b) Urbach energy map of a thin film over the same region of the sample. Scalebars are 1 μm. c) 2-dimensional kernel density estimation plot of Urbach energy versus PLQE showing anticorrelation between the two properties (Spearman's r-value = -0.44, p-value << 0.0001). d) Spatially averaged PL spectra of the highest (>80$^{th}$ percentile, blue) and darkest (<20$^{th}$ percentile, light gray) PLQE regions highlighting a red-shifted shoulder in the most emissive regions.

To unveil the nature of the nanoscale landscape and how it influences the optoelectronic performance, we probe the chemical and structural properties locally on the same scan area as the luminescence measurements by performing simultaneous synchrotron-based nano-X-Ray Fluoescence (nXRF) and nano-X-Ray Diffraction (nXRD) measurements, respectively, with a spatial resolution of ~ 50 nm (see Methods including discussion of beam sensitivity). The spatial variation in chemistry is extracted by tracking the characteristic Pb $L_\alpha$, I $L_\alpha$ and Br $K_\alpha$ lines in nXRF (see spectrum in Supplementary Fig. 7), and we use ratios of the halide and lead XRF signal intensities to assess compositional variations independent of thickness changes (Supplementary Fig. 8). While the iodine to lead (I:Pb) ratio does not vary much across the region of interest (Supplementary Fig. 9), there are larger variations in the bromine-to-lead (Br:Pb) ratio of approximately ±10% on a length scale of hundreds of nanometres (Figure 3a, see Methods). The nXRD measurements obtained simultaneously on the same scan area reveal clusters of structural variations across the films on a similar length scale to the nXRF ascertained by extracting the cubic lattice parameter at each point from the nXRD pattern, with variations of the extracted lattice parameter from the 210 plane within clusters of up to ±0.5% (Supplementary Fig. 10). We note that each individual nXRD peak is only observable on a subset of the scan region because diffraction information is only observed when the Bragg condition is met at each probe point. Using the simultaneous measurement of structure and composition, we decouple the underlying contribution to the structural variations from chemical (halide) variations by extracting the expected cubic lattice parameter of the perovskite from the Br:Pb ratio shown in Figure 3a using Vegard's law for the $FA_{0.79}MA_{0.15}Cs_{0.05}Pb(I_xBr_{1-x})_3$ material family ($A_{XRF}$), and compare to the lattice parameter extracted from the nXRD measurements ($A_{XRD}$) (see Methods). The measured $A_{XRD}$ shows a ~7 times wider spread of lattice constants than we would expect given the chemical variation $A_{XRF}$ alone (Figure 3c); we interpret this residual variation as nanoscale strain variations in the 210 family of planes of ±0.5% (Figure 3b) and up to ±1% in other planes (Supplementary Fig. 10). Thus, we reveal strain clusters on the length scale of hundreds of nanometres. Intriguingly, we find the spatial mean of the map is net tensile strained, which has been previously shown to be caused by a mismatch in the thermal expansion coefficient between the perovskite and the substrate[48]. We spatially correlate these strain variations with the $E_U$ and find that there is a very weak but statistically significant positive correlation (Spearman's r value = 0.07, p-value = 4×10⁻³) between strain and $E_U$, as shown in Figure 3d. Importantly, this correlation cannot adequately account for the spatial variation in performance observed. These data suggest that the variance in the optoelectronic quality and properties of the sample cannot be explained by nanoscale strain variations along the planes we have measured in these mixed composition systems.

We now consider how the chemical composition impacts the optoelectronic properties by overlaying the regions with higher and lower Br content (>80[th] and <20[th] percentile of Br:Pb ratio, respectively) over the $E_U$ maps in Figures 3e and 3f, respectively. Interestingly, the areas showing higher Br content correlate well with those of lower Urbach energy, and vice versa with lower Br and higher Urbach regions. This is highlighted in Figure 3g, where the histograms for the two populations show a statistically significant shift between the two populations (two-sample Kolmogorov-Smirnov test, p<<0.0001, see Methods). PLQE maps of the same area overlaid with higher and lower Br areas (Supplementary Fig. 11) also show a strong and statistically significant correlation between high PLQE and higher Br content (p<<0.0001; Figure 3h). This striking observation seems to apparently contradict the observation of the local red-shifted PL peak (Figure 2) in the higher Br region – a PL feature typical of more I-rich perovskite compositions.

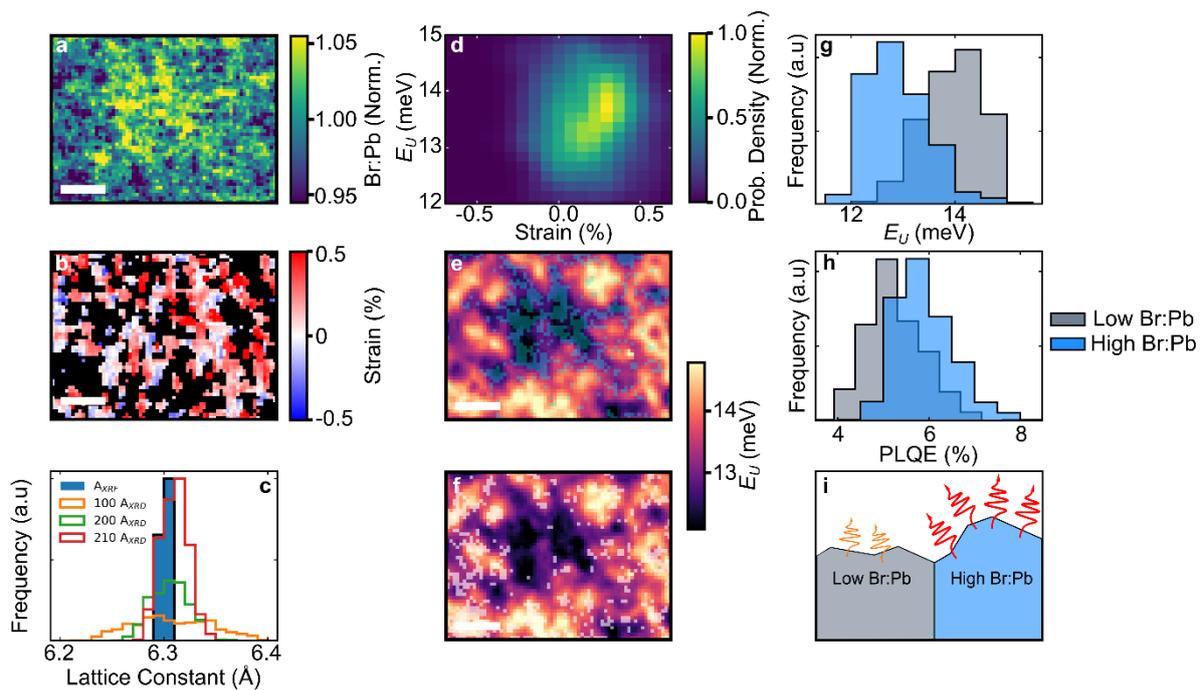

**Figure 3**: **Spatial relationships between halide composition, structural and optoelectronic variations in $FA_{0.79}MA_{0.16}Cs_{0.05}Pb(I_{0.83}Br_{0.17})_3$ perovskite films.** a) Normalised Br:Pb ratio map of a thin film. b) Strain map of the 210 family of planes extracted from nXRF and nXRD of the same area as in a). c) Histograms of the lattice constants extracted from nXRF ($A_{XRF}$) and from nXRD ($A_{XRD}$) from the 100, 200 and 210 peaks. d) 2-dimensional kernel density estimation plot of Urbach energy versus 210 strain showing a weak positive correlation between the two properties (Spearman's r value = 0.07, p-value = 4 ×$10^{-3}$). e) Urbach energy maps overlaid with the regions of highest Br (>80[th] percentile, light blue) and f) lowest Br (<20[th] percentile, light gray) content, respectively. g) Histograms of the Urbach energies and h) PLQEs in the highest Br (>80[th] percentile, blue) and lowest Br content regions (<20[th] percentile, light gray). The populations are statistically different in both cases as determined by two sample Kolmogorov-Smirnov testing (p<<0.0001). i) Schematic highlighting emission behaviour in high and low Br regions. All scalebars are 2 µm.

These measurements reveal that the higher Br:Pb regions exhibit the highest PLQE and the lowest electronic disorder, even in spite of the presence of a low-energy emission shoulder, which is itself suggestive of local chemical disorder (Figure 3i). By contrast, the regions with low Br content have the lowest PLQE, the highest electronic disorder ($E_U$) but simultaneously the lowest apparent chemical disorder based on their narrow PL spectral linewidth. Furthermore, the strain variations have only a weak influence on the properties, further emphasizing the key role played by the local chemical variations. This is remarkable given strain of such magnitude (~1 %) would dramatically influence the optoelectronic properties of a range of 2D materials[49], and would produce dislocation defects that would compromise performance of III-V semiconductor bulk and nanocrystal devices[50,51]. We find that these results hold across other regions and sample batches investigated (Supplementary Fig. 12). We note that the relationship between the chemistry and optoelectronic properties is not an intrinsic material property but can also be modulated with film formation conditions in which the segregation is exacerbated (Supplementary Fig. 13). We attribute the regions with lower Br:Pb fractions that have high electronic disorder and low PLQE to regions where deep trap clusters act as non-radiative recombination sites, as reported recently[22,52]; a clear correlation between clusters with high sub-gap trap density, ascertained through local photo-emission microscopy (PEEM), and lower Br:Pb content, through nXRF measurements, confirms this proposition (Supplementary Fig. 14). Therefore, these multimodal measurements reveal the important features of nanoscale heterogeneity. Specfically, there are two primary, distinct types of sites that influence local performance: sites with high Br content with low electronic disorder but an intriguing low energy emission shoulder corresponding to hot spots of high performance, and sites with low Br content containing deep trap states associated with non-radiative recombination and electronic disorder (Figure 3i).

In order to understand the nanoscale charge carrier recombination competition between these sites and the nature of the low bandgap recombination within a nominally Br-rich region, we perform transient absorption microscopy (TAM) measurements[53] spatially correlated with nXRF measurements (Supplementary Fig. 15). The wide-field TAM methodology allows us to simultaneously monitor the temporally and spectrally resolved differential transmittance signal ($\Delta T/T$) for each spatial point (see Methods). Here, we focus on characterising the ground state bleaching (GSB) band of the transient absorption spectra, which are positive changes in $\Delta T/T$ stemming from a bleaching of the transitions between the valence and conduction bands [54]. Monitoring the GSB signal allows us to monitor the extent of the excited carrier population as a function of energy, therefore enabling us to infer the size of the bandgap carriers experience as a function of space and time. In Figure 4a, we show the Br:Pb ratio of an nXRF line-scan correlated with the PL profile reconfirming the earlier trend of increased steady-state PL intensity in regions of higher Br:Pb ratio on the region of interest for TAM. We then plot the centre-of-mass energy of GSB band as a function of time in Figure 4b. We further highlight areas rich in Br (blue dashed line) and areas deficient in Br (grey dashed line) and compare the local energetic evolutions of those two distinct regions (Supplementary Fig. 16). In the Br-rich region, photoexcited carriers initially occupy higher energy states (~1.66 eV) relative to the Br-poor region (~1.64 eV). The GSB band in the Br-rich region then rapidly red-shifts towards 1.63 eV on the timescale of hundreds of picoseconds, which is evident in the local TA spectra snapshots shown in Figure 4c. In contrast, the Br-poor region behaves in a remarkably different manner (Figure 4d, grey dashed line) with the GSB centre-of-mass energy remaining constant at ~1.64 eV throughout the 500-ps window, consistent with carrier trapping. The GSB centre-of-mass, energy shift profile in Figure 4e highlights this considerable disparity between the two regions. The continuous red shift of the peak in the Br-rich region indicates that charge carriers are funneled to slightly lower energy (~ 20 meV) domains, consistent with the lower-energy emission behaviour observed in the steady-state hyperspectral PL measurements (cf. Figure 2d). Moreover, larger local energy shifts are linked to sample areas with higher PL intensity (Supplementary Fig. 17). Interestingly, the local TA signal from the Br-poor region shows a higher energy peak at ~1.65 eV that drops in intensity and a low energy shoulder at ~1.56 eV that remains constant with time (Figure 4d). This suggests that in the Br-rich regions there are local inclusions of lower bandgap material onto which the carriers transfer, whereas this energetic transfer is not observed in the I-rich regions.

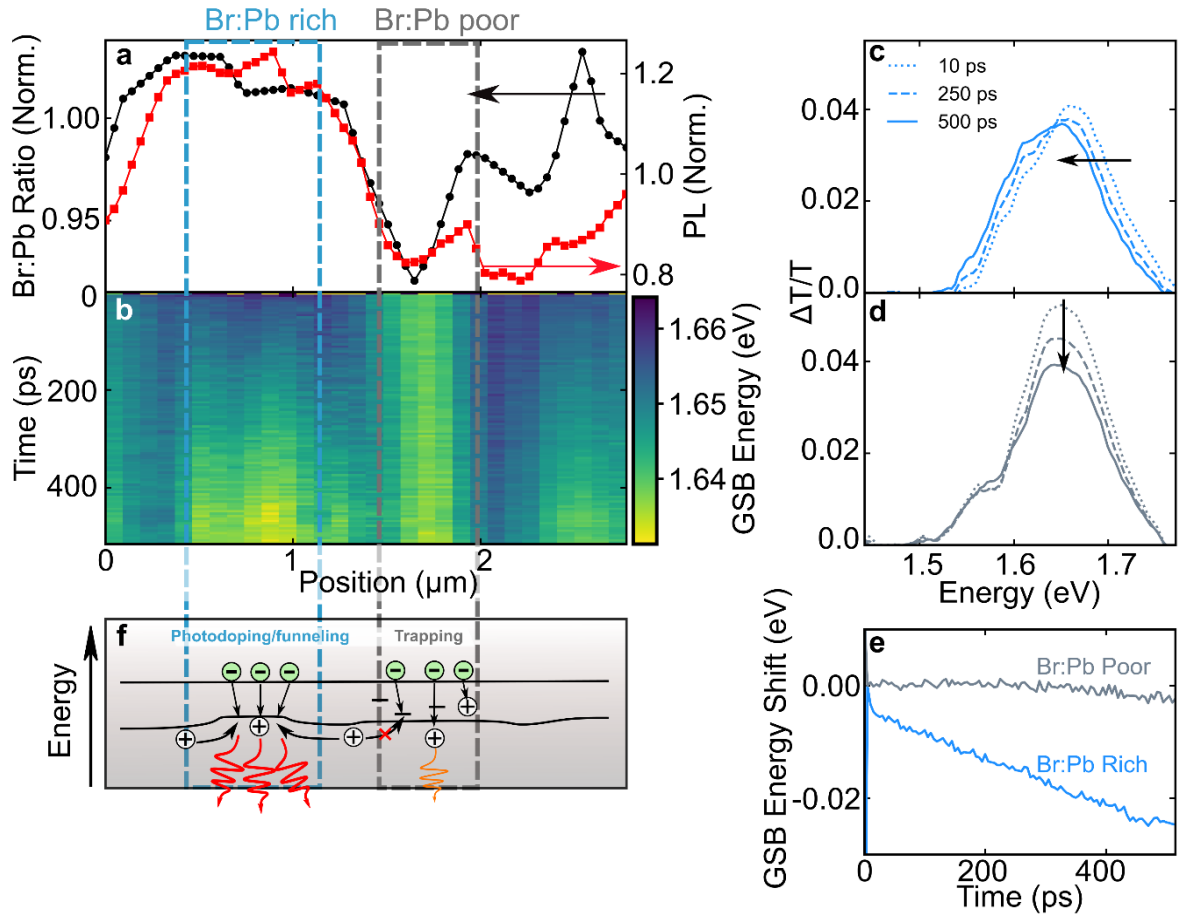

**Figure 4**: **Transient absorption microscopy of FA$_{0.79}$MA$_{0.16}$Cs$_{0.05}$Pb(I$_{0.83}$Br$_{0.17}$)$_3$ perovskite films correlated with local chemical mapping.** a) Linescan of the Br:Pb nXRF intensity ratio (black dots) and normalised PL intensity (red squares) of the region used for correlation. b) TAM linescan showing the centre-of-mass value of the GSB as a function of time after excitation. The Br-rich and Br-poor regions are denoted with blue and grey rectangles respectively. c) The GSB TA spectra of the Br-rich and d) Br-poor regions at different delay times. e) The relative GSB energy centre of mass shift profile of the two regions. f) Schematic of the model proposed in this work showing carriers funneling to highly emissive low bandgap regions in Br-rich regions, outcompeting carrier trapping in Br-poor regions.

Taking all of the microscopic observations together, we elucidate a global picture of the nanoscale landscape and how it endows enhanced defect tolerance on the seemingly highly disordered perovskite material. Mixed-composition perovskites exhibit a spatially heterogeneous landscape of chemical composition, structural variation in the form of strain and trap state density imprinted onto the sample upon film formation (Figure 4f) that dictates where charges accumulate and recombine. Large, micrometer-size bulk regions slightly rich in bromide also contain low bandgap inclusions (below detection limits for the compositional maps) that are of very high optoelectronic quality. Photons absorbed in these areas excite charge carriers in the Br-rich bulk that readily funnel down a shallow energy gradient to these lower energy, more ordered inclusions. This provides, for the first time, a complete nanoscale visualisation and hence a mechanistic understanding of the macroscopic observations of photo-doping [34] in which holes are proposed to locally accumulate at lower bandgap regions. Such hole accumulation in local, high-quality regions leads to high carrier densities and extremely efficient radiative recombination (cf. Figure 1f). These regions are comprised of multiple morphological grains and are up to the order of micrometers (see Figure 3 and Figure 4a) in size, leading to an effective, large capture cross-section for harvesting charges in the funnel. By contrast, the deep trap clusters with low electronic quality that trap holes[22] are tens to hundreds nanometers in size and uniquely associated with Br-poor regions in these mixed halide compositions (Supplementary Fig. 14). Therefore, the larger capture radius of the shallow energy funnel set up by the compositional disorder effectively outcompetes the smaller capture cross-section of carrier traps associated with electronic disorder, leading to high luminescence yields and strong overall device performance (Figure 4f). By contrast, single-halide iodide-only analogues have an order of magnitude lower luminescence yield as they do not exhibit such carrier funneling (cf. Supplementary Fig. 6). The presence of nanoscale, complex strain variations, surpisingly, do not strongly influence the optoelectronic behaviour, providing further evidence that the local chemistry dictates the local performance and further highlights the defect tolerance of these materials.

Our study has significant implications for the fundamental understanding of defect tolerance in these materials and the design of halide perovskite solar cells, in particular for tandem cells. Using a novel suite of multimodal microscopy techniques, we unveil the remarkably complex energetic landscape that charge carriers must navigate in halide perovskites. We provide the first nanoscale picture of how this energetic landscape influences photodoping, carrier recombination and trapping. We find that the pursuit of homogeneous chemical compositions is not necessarily the best way to maximize the performance of this family of semiconductors, at least while the material still possesses deep trap clusters that lower device performance from the radiative limits. The existence of mixed Br and I samples induces the formation of beneficial local heterostructures that confer enhanced defect tolerance to these materials. In these regions, charge-carrier photogeneration and radiative recombination occurs through a rapid wide-to-narrow bandgap funneling process, more efficient than in the chemically homogeneous counterparts. The chemical disorder in these regions allows the effective capture of diffusing carriers over micron-length scales to lead to radiative recombination, outcompeting the capture of carriers in more electronically disordered and trap-rich regions, thus resulting in strong luminescence and performance. Furthermore, nanoscale strain variations even of order 0.5% do not strongly impact performance, providing further confirmation of the purported defect tolerance and the dominance of the chemical properties. Therefore, this effect can be interpreted as a form of spatial enhanced defect tolerance serendipitously occurring in these mixed composition materials that have been optimised through empirical device observations. The beneficial chemical disorder in these perovskite compositions is key to their impressive device performance, where the small penalty to pay in QFLS (voltage) of ~10-20 meV due to accumulation of carriers in slightly lower bandgap performance hot spots (with low energetic disorder) outweighs the potentially larger loss if carriers find regions of high energetic disorder such as deep trap clusters. This process is in part analogous to energy funneling in mixed thickness, quasi-2D perovskites used for high efficiency LEDs [55], but here the chemical gradients and thus energy losses are far more subtle yet as effective. Nevertheless, such heterogeneity may eventually become detrimental in solar cells as we approach the radiative limit in which deep trap states are at negligible levels where defect tolerance is no longer necessary. This multimodal methodology represents a large step forward in the understanding of the fundamental processes and nanoscale landscape governing these intriguing defect-tolerant materials and is widely applicable to the study of other emerging semiconductors at the nanoscale.

**Acknowledgements**

K.F. acknowledges a George and Lilian Schiff Studentship, Winton Studentship, the Engineering and Physical Sciences Research Council (EPSRC) studentship, Cambridge Trust Scholarship, and Robert Gardiner Scholarship. M.A. acknowledges funding from the Marie Skłodowska-Curie actions (grant agreement No. 841386) under the European Union's Horizon 2020 research and innovation programme. S.M. acknowledges an EPSRC studentship. T.A.S.D. acknowledges a National University of Ireland Travelling Studentship. The authors acknowledge the Diamond Light Source (Didcot, Oxfordshire, UK) for providing beamtime at the I14 Hard X-ray Nanoprobe facility through proposals sp19023 and sp20420. Additionally the authors thank Julia E. Parker and Paul D. Quinn for their assistance and guidance on the I14 beamline and with data handling. S.D.S. acknowledges the Royal Society and Tata Group (UF150033). The work has received funding from the European Research Council under the European Union's Horizon 2020 research and innovation programme (HYPERION - grant agreement no. 756962, SOLARX - grant agreement no. 758826).Y.H-C. thanks the Cambridge Trust for a studentship. The authors acknowledge the EPSRC (EP/R023980/1, EP/M006360/1) and the Winton Programme for the Physics of Sustainablity for funding. A.J.W. and K.M.D. acknowledge that this work was supported by the Femtosecond Spectroscopy Unit of the Okinawa Institute of Science and Technology Graduate University and JSPS Kakenhi Grant Number JP19K05637. The authors acknowledge the support for this work from the Imaging Section and Engineering Support Section of the Okinawa Institute of Science and Technology Graduate University.

**Competing Interests**

SDS is a cofounder of Swift Solar.

**Data Availability Statement**

The data that support the findings of this study are available in the University of Cambridge Apollo repository at [DOI to be generated and added]

**Materials and Methods**

Film and Device Fabrication

ITO substrates (12 x12 mm$^2$, Kintec) and glass cover slips were cleaned with the following steps: deionized water with 1.5% v/v solution of Decon 90 detergent, deionized water, acetone, and finally isopropanol for 15 minutes per step in an ultrasonic bath. ITO and glass substrates were further cleaned by UV-Ozone for 15 minutes and transferred to the glovebox immediately. SiN X-ray transparent grids (Norcada, product number NX7100C) were cleaned in an oxygen plasma cleaner for 5 minutes before immediately transferring to the glovebox. PTAA (2mg/ml, Sigma) in anhydrous chlorobenzene (Sigma) was prepared and dropped on the middle of the substrate and then spin coated at 5000 rpm (1000 rpm/s ramp) for 30 seconds. PTAA films were annealed at 100° C for 10 minutes. To improve the surface wetting and passivate the PTAA layer, a PFN-P2 layer (0.3 mg/ml, dissolved in anhydrous methanol) was spun on the top of the PTAA at 5000 rpm (2500 rpm/s ramp) for 30 seconds and annealed at 100 degree for 10 mins. For perovskite solution preparation, a composition of $(Cs_{0.05}FA_{0.79}MA_{0.16})Pb(I_{0.83}Br_{0.17})_3$ was used, containing formamidinium iodide (FAI) (1 M,

Greatcell Solar), methylammonium bromide (MABr) (0.2 M, Greatcell Solar), PbI$_2$ (1.1 M, TCI), PbBr$_2$ (0.2 M, TCI) dissolved in anhydrous dimethylformamide/dimethyl sulfoxide (DMF:DMSO 4:1 (v:v), Sigma). 5% CsI (Sigma) dissolved in DMSO (1.5 M) was then added to the precursor solution. To prepare the perovskite film, 50 µl of prepared solution (20 µl for the SiN grids) was spread onto the substrate and spun in a two-step spinning process: 1000 rpm for 10 s and 6000 rpm for 20 s. During the second spinning, 100 ul of Chlorobenzene was dropped in the middle of film 5 s before the end of the process. After spinning, the substrates were moved on the hotplate for an annealing step with 100 degree for 1h. After cooling down the perovskite to room temperature, PCBM (20 mg/ml, pre-heated solution at 75 degree for 3hours) was spread on the perovskite film with a spin-coating process of 1200 rpm for 30 s. The as-prepared BCP (0.5 mg/ml in IPA, Alfa) solution was dynamically spun onto the PCBM layer with a speed of 2000 rpm for 30 sec. To finish the full device, the Ag metal layer was thermally evaporated on the samples to achieve the desired thickness. At each step, the SiN substrates were transferred onto the hotplate on a glass slide to avoid thermal shock of the thin window.

Solar Cell Device Characterisation

Current-voltage (J-V) curve measurements were performed using a xenon lamp (Abet Sun 2000 Solar Simulators, AAB class) at 1 sun intensity illumination (100 mW/cm$^2$, AM 1.5G), calibrated with a reference Silicon diode (no window filter). The spectrum mismatch factor was calculated and applied onto the solar simulator power. We recorded the device performance with a Keithley 2636A, source meter, controlled by a home-built LabVIEW program. We started the J-V measurements from a direction of 1.2 to -0.1 V (reverse scan) to -0.1 to 1.2 V (forward scan) with a step size of 20 mV and a delay time of 100 ms. The active area of solar cell was 4.5 mm$^2$, defined by the overlapping area between Ag and ITO. All the devices were being measured in air with any encapsulation.

Hyperspectral Microscope Characterisation

Wide field, hyperspectral microscopy measurements were carried out using a Photon Etc. IMA system. 100x air and oil, chromatic aberration corrected objective lenses from Olympus (MPLFN and MPLAPON) were used for all measurements. All samples were stored in a nitrogen filled glovebox until immediately before measurement to mitigate oxygen and humidity related transient behaviour. A 405 nm continuous wave laser was used for luminescence excitation. 50 W halogen lamps were used for transmission and reflection measurements. The excitation laser was filtered by a dichroic mirror. The lamp light used for reflection measurements travels through the objective to the sample. The lamp used for transmission measurements was focused on the sample by a condenser lens from below the sample and is collected by the objective lens. The emitted/transmitted/reflected light from the sample was incident on a volume Bragg grating, which splits the light spectrally onto a CCD camera. The detector was a 1040×1392 resolution silicon CCD camera that is kept at 0 °C with a thermoelectric cooler and has an operational wavelength range of 400-1000 nm. By scanning the angle of the grating relative to the incident light, the spectrum of light coming from each point on the sample could be obtained.

For calibration of the system to extract the absolute number of photons at each point, a two step process was used for each objective lens used. Firstly, a calibrated white light lamp from Ocean Optics was coupled into an integrating sphere. The objective lens was also coupled into the integrating sphere. Comparing the measured spectrum of the lamp at each point to the calibrated spectrum gives the relative sensitivity of the system both spectrally and spatially. Secondly, a 657 nm laser was coupled directly to the microscope by an optical fibre. The power of the laser was measured precisely at the output of the fibre using a power meter before coupling to the objective lens. Measuring the laser on the system allows direct conversion between number of counts and photons at this wavelength. Combining this absolute calibration with the relative calibration from the calibrated white light lamp and integrating sphere allows absolute calibration across the spectrum at each point of the sample.

To determine local absolute reflectance, the macroscopic reflectance spectrum of a calibration mirror was measured. Before each hyperspectral measurement of a given perovskite sample, a hyperspectral measurement of the calibration mirror was also measured. A spatial median filter was applied to this data to reduce influence from local imperfections in the mirror. The reflected spectrum from the mirror at each point was divided by the macroscopic reflectance spectrum to obtain the calculated incident lamp spectrum. Once a sample had been measured in reflection, the data was divided by this lamp spectrum in order to obtain the absolute reflectance of the sample. An equivalent process was also performed for transmittance using a reference glass cover slip. This is summarised in the following equation where, $R(E,x,y)$ is the reflectance from the sample at position $(x,y)$ and energy E, $I_{sample}(E,x,y)$ is the intensity of light reflected from the surface of the sample at point $(x,y)$ and at an Energy E in the hyperspectral microscope, $I_{mirror}(E,x,y)$ is the same for the reference mirror, and $R_{mirror}(E)$ is the measured reflectance of the mirror at an energy E:

$$R(E,x,y) = \frac{I_{sample}(E,x,y)}{I_{Mirror}(E,x,y)/R_{Mirror}(E)}$$

Suns-Intensity Calculation

In order to calculate the equivalent number of suns for a given monochromatic excitation of a given power, we used an interpolated AM 1.5G spectrum and converted from units of spectral irradiance (W m$^{-2}$ nm$^{-1}$) to photons m$^{-2}$ nm$^{-1}$ s$^{-1}$ by dividing by the photon energy at each point of the spectrum. Then the spectrum was integrated over wavelength from 300 nm to the bandgap energy of the material (in this case ~765 nm) to obtain the total flux of above bandgap photons, photons m$^{-2}$ s$^{-1}$. This flux was compared to the photon flux incident on the sample from the monochromatic excitation to obtain the equivalent number of suns.

Quasi-Fermi Level Splitting Extraction

Quasi-Fermi level splitting (QFLS) can be extracted from absolute PL intensity based on the generalised Planck law [56,57]. The intensity of PL, $I_{PL}$ can be thought of as the product of the photon density of states ($\rho(E)$), a Bose-Einstein occupation function ($f_{BE}(E)$) and the absorptance spectrum of the material ($a(E)$)

$$I_{PL}(E) = \rho(E) \times f_{BE}(E) \times a(E)$$

$$\rho(E) = \frac{2\pi E^2}{h^3 c^2}$$

$$f_{BE}(E) = \frac{1}{e^{(E-\Delta\mu)/kT} - 1}$$

Where E is energy, h is Planck's constant, c is the speed of light, k is Boltzmann's constant and $\Delta\mu$ is the QFLS. The most common approach for QFLS extraction in the literature is first to approximate the Bose-Einstein occupation function as a Boltzmann distribution, one then takes the logarithm of both sides, and assuming the absorptance spectrum is approximately constant at the high energy side of the PL spectrum, the QFLS spectrum can be extracted with a linear fit.[20,58-60]

$$I_{PL}(E) = \frac{2\pi E^2}{h^3 c^2} \times a(E) \times exp\left(-\frac{E - \Delta\mu}{kT}\right)$$

$$ln\left(\frac{I_{PL}(E)}{E^2}\right) = ln\left(\frac{2\pi}{h^3 c^2}\right) - \frac{E}{kT} + \frac{\Delta\mu}{kT}$$

An alternative approach where a model for the absorptance is also included and the entire PL peak can be fit was described by Katahara and Hillhouse [38,61]. The model for the absorption coefficient is a convolution of the above bandgap density of states which varies with for a conventional 3D semiconductor as:

$$\alpha = \alpha_0 \sqrt{E - E_g}$$

where α₀ is a parameter that depends on the oscillator strength of the material and $E_g$ is the bandgap, and a below bandgap density of states:

$$\alpha \propto \exp\left(\frac{E_g - E}{\gamma}\right)^\theta$$

where γ is the characteristic energy broadening of the low energy tail and θ determines the form of the exponential tail. When θ=1, γ is the Urbach energy. For values of θ greater than 1, this can be interpreted as longer range disorder in the material [62]. The combination of the two can be approximated as:

$$\alpha = \alpha_0 \sqrt{\gamma} G\left(\frac{E - E_g}{\gamma}, \theta\right)$$

Where $G(\frac{E-E_g}{\gamma}, \theta)$ is a function where the convolution integral has been explicitly evaluated and lookup tables provided by Braly et al [61]. Assuming that the absorption coefficient is occupation independent at low excitation densities, one can then write the overall intensity of PL as:

$$I_{PL}(E) = \frac{2\pi E^2}{h^3 c^2} \times \left(1 - \exp\left(-\alpha_0 d \sqrt{\gamma} G\left(\frac{E - E_g}{\gamma}, \theta\right)\right)\right) \times \exp\left(-\frac{E - \Delta\mu}{kT}\right)$$

Where d is the thickness of the film. The model has been successfully used on perovskites amongst other systems [63]. The $\alpha_0 d$ product was set to be 10 as is common for perovskite materials and has little impact on the fitted values of other parameters [61]. The temperature was set to the temperature of the room, generally 300 K. The remaining parameters to be fit are $\gamma$, $E_g$, $\theta$ and $\Delta\mu$. The data was fit using the Levenberg-Marquardt, non-linear least squares fitting algorithm implemented in Python (see Figure S1). In order to set initial guess parameters, the average PL spectra of the region of interest was fit and the output manually inspected before automatic fitting of the individual pixels was performed.

Urbach Energy Extraction

The Urbach energy was extracted from the red, low energy portion of the PL spectrum. In order to be directly comparable to other works, the standard Urbach energy was reported rather than the more complicated coupled energy broadening and exponential powers mentioned above. The generalised Planck law was rearranged for absorptance:

$$a(E) = I_{PL}(E) \times \frac{h^3 c^2}{2\pi E^2} \times \exp\left(\frac{E - \Delta\mu}{kT}\right)$$

The absorptance is related to the absorption coefficient as mentioned previously as:

$$a(E) = 1 - \exp(-\alpha d)$$

For small values of $\alpha d$ below the bandgap, one can perform a Maclaurin expansion of the exponential which is to first order:

$$a(E) \approx 1 - (1 - \alpha d) = \alpha d$$

The absorptance therefore follows the same functional form as the absorption coefficient for small arguments. We write the absorption coefficient in the Urbach tail as:

$$\alpha = \alpha_0 \exp\left(\frac{E_g - E}{E_U}\right)$$

Therefore, if we take the log of both sides:

$$\ln(\alpha) = \ln\left(I_{PL}(E) \times \frac{h^3 c^2}{2\pi E^2} \times exp\left(\frac{E - \Delta\mu}{kT}\right)\right)$$

Finding the inverse slope of this function using a Huber loss function, linear regression to eliminate effects of noisy outliers of data when mass fitting entire maps at the red edge of the PL gives the Urbach energy as shown in Figure S2.

Local Photoluminescence Quantum Efficiency (PLQE) Extraction

In order to extract the local PLQE, first the flux of absorbed photons from the excitation laser must be determined. First, the power of the 405 nm laser at the objective lens of the microscope was measured using a power meter. The spot size of the laser was then measured to determine incident laser intensity. Then the local reflectance (R) and transmittance (T) of the sample was measured. At short wavelengths < 500 nm, the transmittance is below our threshold of detection and is set to be zero. Then absortance (A) was assumed to be 1-R in this wavelength range. The power of the lamps and the sensitivity of the CCD and volume Bragg grating at 405 nm prevented accurate extraction of A at 405 nm, so the absorptance averaged from 450-490 nm was used instead for each point. The intensity of the incoming excitation times the probability of absorption gives the number of photons absorbed at each point per second. Integrating the absolute PL spectrum over energy gives the total number of emitted photons per second. The ratio of these two values gives the local PLQE.

Bulk X-ray diffraction (XRD) measurements

Bulk XRD measurements were performed with a Bruker D8 ADVANCE system with a Copper focus X-ray tube ($K_\alpha$: 1.54 Å, 40 kV). The scan range of 2θ was from 3° to 40° with a step size of 0.01° and a delay time of 0.15 s, corresponding to a q range of ~0.2-2.8 Å$^{-1}$. The sample was measured in air.

Synchrotron Nano X-ray Diffraction(nXRD) and Nano X-ray Fluorescence (nXRF)

Measurements were performed on the i14 Hard X-ray Nanoprobe beamline at Diamond Light Source Ltd., Didcot, UK. Samples were stored in nitrogen prior to measurements. The full experimental setup has been described elsewhere [64]. Briefly, X-rays from an undulator source are monochromated to produce a 19 keV, monochromatic X-ray beam. The beam travels along a 186 metre long channel, along the path, the beam is focused with Kirkpatrick Baez mirrors to produce a X-ray with FWHM of ~50 nm at the focus. A sample is placed on a raster scanning stage and is raster scanned across this focal point. The energy resolved nXRF signal is collected with a 4 element silicon drift detector in a reflection geometry. The diffracted X-rays are collected in transmission with an Excalibur 3M detector consisting of 3 Medipix 2048 x 512 pixel arrays. For measurements involving nXRD/nXRF correlations, samples were deposited on X-Ray transparent, SiN windows from Norcada (product number NX7100C). The step size for the measurements was 50 nm and the dwell time was 0.75 s per point. All samples were stored in a nitrogen filled glovebox before measurement, and held under a nitrogen flow during measurements. Multidimensional nXRD and nXRF datasets were handled and analysed using the open source Python package Hyperspy [65]. We note that we cannot rule out some damage to the perovskite following measurements owing to the highly focused X-ray nanoprobe. However, given that the acquired nXRD patterns (acquired simultaneously to the nXRF) index to the expected cubic $FAPbI_3$ model (without any impurity phases) and previous work has shown that the nXRF signal is more beam stable than the nXRD [66], we assert that our measured maps and conclusions are not influenced by beam damage. Furthermore, the strong agreement between our diffraction/chemical based conclusions from scanning probe measurements with metrics obtained from optical measurements further support our conclusions. We focus on the Br:Pb ratio rather than I:Pb as the Br:Pb is a more sensitive probe of the chemistry in this system as the iodine occupies a much larger stoichiometric fraction than bromine, such that a change of a few atomic percent at the X-site is a much larger relative change for the bromine than iodine.

Chemistry Normalised Strain Extraction

The simultaneous measurement of nXRF and nXRD on the same region allows us to remove the contribution of local chemistry and just focus on variations in the lattice parameter due to strain. We assume a Vegard's law holds for the $FA_{0.79}MA_{0.15}Cs_{0.05}Pb(I_xBr_{1-x})_3$ family over the range we look at in these samples. The lattice parameter of the pure iodide $FA_{0.79}MA_{0.15}Cs_{0.05}PbI_3$ is ~6.36 Å and the nominal composition of the sample here,

FA$_{0.79}$MA$_{0.15}$Cs$_{0.05}$Pb(I$_{0.83}$Br$_{0.17}$)$_3$ has a lattice parameter of 6.30 Å. We look at the Br:Pb ratio, we assume that the composition in our solution is the same as that in the film, and we set the mean value of the Br:Pb ratio to correspond to this 17% Br composition. Variations about this mean value are converted to lattice parameters (A$_{XRF}$) by inserting the extracted composition into Vegard's law as mentioned above. Simultaneously, the assumed cubic lattice parameter is extracted at each point from the various nXRD peaks using:

$$\frac{1}{d^2} = \frac{h^2 + k^2 + l^2}{A_{XRD}^2}$$

Where d is the plane spacing extracting from the q value of the peak, h, k and l are the Miller indices of the plane and A$_{XRD}$ is the cubic lattice parameter extracted from XRD. Percentage strain is then calculated as:

$$Strain(\%) = \left(\frac{A_{XRD}}{A_{XRF}} - 1\right) * 100$$

Image Registration for Correlations

A suspension of pseudo-planar Au nanoparticles with distinctive shapes (hexagons and triangles) were used as fiducial markers as per our previous works [22,31]. The gold nanoparticles were dispersed in chlorobenzene (Sigma) and spun onto the perovskite film. For image alignment, optical and nXRF images of the same region were overlaid in the open source image editing software GIMP. The optical images were transformed using the unified transform tool in GIMP which includes all of the operations associated with an affine transformation (translation, scaling, shearing and rotation) along with a perspective transform to account for distortions in the detectors as well as any relative sample tilt between the measurements. The transform was applied until the Au fiducial markers were overlaid and simultaneously features from the Pb nXRF signal matched those from the reflectance images

Statistical Analysis

Differences in the populations of lower and higher Br:Pb regions were calculated using a two sample Kolmogorov-Smirnov test implemented in the Scipy python library[67] using the ks_2amp functionality. Output p values <<0.0001 were interpreted as the null hypothesis of the two samples being part of the same population being rejected and a statistically significant difference between them.

Correlations between data were tested using the Spearman's r-value which tests whether the relationship between two datasets can be described with a monotonic function. This was also implemented in the Scipy python library using the spearmanr function. P-values << 0.0001 were interpreted as a statistically significant relationship between the two variables.

Transient Absorption Microscopy (TAM)

A Yb:KGW laser system (Pharos, Light Conversion) provided the fundamental output beam of 200 fs, 30 µJ pulses at 1,030 nm with 200 kHz repetition rate. The output beam was divided by a beam splitter and seeded two broadband white light continuum (WLC) stages. The WLC for the probe beam was generated in a 3 mm YAG crystal, covering the wavelength range from 650 to 950 nm selected by a fused-silica prism-based spectral filter. In contrast, the WLC for pump beam was generated in a 3 mm sapphire crystal to extend the WLC to 500 nm, and short-pass filtered at 650 nm (FESH650, Thorlabs). A set of chirped mirrors (pump, 109811, Layertec; probe, DCM9, Venteon) and a pair of fused silica wedges (Layertec) were used to compress the pulses to 9.2 fs (pump) and 6.8 fs (probe). The time resolution was then verified by second-harmonic generation frequency-resolved optical gating (FROG). A closed-loop piezo translation stage (M-ILS100HA, Newport) was used to delay the probe with respect to the pump. Both pump and probe beams were loosely focused onto the sample with a focal spot size of 30 µm by a dispersion-free concave mirror. This allowed even-illumination over the region of interest. The transmitted probe pulse was then collected by an oil immersion objective (×100, an effective numerical aperture, NA, of 1.1) and sent to an EMCCD camera (Rolera Thunder, QImaging). The total magnification of the imaging system was ×288. The scattered pump light was rejected by a 650 nm long-pass filter (FEL650, Thorlabs) inserted in front of the camera. Differential imaging was achieved by modulating the pump beam at 30 Hz by a mechanical chopper. In the current experimental scheme, we adopted a slit-prism combination and utilized the configuration in order to achieve simultaneous spatial, temporal and spectral imaging. First, the dimension of the collimated probe beam was reduced by inserting a slit in an image plane. The transmitted 1D probe beam is then dispersed by a prism, enabling us to simultaneously obtain 2D spatial and spectral map as a function of time [68].

Photoemission Electron Microscopy (PEEM)

The fundamental output of a 4 MHz, 650 nJ Ti:sapphire oscillator (FemtoLasers XL:650) delivering 45 fs pulses at 800 nm, was used to generate UV pulses for PEEM measurements. The 4.65 eV UV pulses are generated as the third harmonic of the 1.55 eV (800 nm) radiation by sum frequency generation of the fundamental and the 3.10 eV (400-nm) second harmonic. The UV beam is focused into the ultra-high-vacuum chamber of the PEEM instrument (SPELEEM, Elmitec GmbH), where it is incident on the sample at a grazing angle of approximately 17°. The probe spot size was approximately 250 μm FWHM. The UV pulses were set to p-polarization. Typical UV pulse fluences used were approximately <100 nJ cm.

**References**

1.  Kim, J. Y., Lee, J.-W., Jung, H. S., Shin, H. & Park, N.-G. High-Efficiency Perovskite Solar Cells. *Chemical Reviews* **120**, 7867-7918, doi:10.1021/acs.chemrev.0c00107 (2020).
2.  Quan, L. N. *et al.* Perovskites for Next-Generation Optical Sources. *Chemical Reviews* **119**, 7444-7477, doi:10.1021/acs.chemrev.9b00107 (2019).
3.  National Renewable Energy Lab. PV Efficiency Chart. (2020).
4.  Al-Ashouri, A. *et al.* Monolithic perovskite/silicon tandem solar cell with >29% efficiency by enhanced hole extraction. *Science* **370**, 1300, doi:10.1126/science.abd4016 (2020).
5.  Jeong, J. *et al.* Pseudo-halide anion engineering for α-FAPbI3 perovskite solar cells. *Nature* **592**, 381-385, doi:10.1038/s41586-021-03406-5 (2021).
6.  Yoo, J. J. *et al.* Efficient perovskite solar cells via improved carrier management. *Nature* **590**, 587-593, doi:10.1038/s41586-021-03285-w (2021).
7.  Yoo, J. J. *et al.* An interface stabilized perovskite solar cell with high stabilized efficiency and low voltage loss. *Energy Environ. Sci.* **12**, 2192-2199, doi:10.1039/C9EE00751B (2019).
8.  Saliba, M. *et al.* Incorporation of rubidium cations into perovskite solar cells improves photovoltaic performance. *Science* **354**, 206-209 (2016).
9.  Saliba, M. *et al.* Cesium-containing triple cation perovskite solar cells: improved stability, reproducibility and high efficiency. *Energy Environ. Sci.* **9**, 1989-1997, doi:10.1039/C5EE03874J (2016).
10. Bai, S. *et al.* Planar perovskite solar cells with long-term stability using ionic liquid additives. *Nature* **571**, 245-250, doi:10.1038/s41586-019-1357-2 (2019).
11. Horantner, M. T. & Snaith, H. J. Predicting and optimising the energy yield of perovskite-on-silicon tandem solar cells under real world conditions. *Energy Environ. Sci.* **10**, 1983-1993, doi:10.1039/C7EE01232B (2017).
12. Hörantner, M. T. *et al.* The Potential of Multijunction Perovskite Solar Cells. *ACS Energy Lett.* **2**, 2506-2513, doi:10.1021/acsenergylett.7b00647 (2017).
13. Abdi-Jalebi, M. *et al.* Maximising and Stabilising Luminescence in Metal Halide Perovskite Device Structures. *Nature* **555**, 497-501, doi:10.1038/nature25989 (2018).
14. Tong, J. *et al.* Carrier lifetimes of >1 μs in Sn-Pb perovskites enable efficient all-perovskite tandem solar cells. *Science* **364**, 475, doi:10.1126/science.aav7911 (2019).
15. Köhnen, E. *et al.* Highly efficient monolithic perovskite silicon tandem solar cells: analyzing the influence of current mismatch on device performance. *Sustainable Energy & Fuels* **3**, 1995-2005, doi:10.1039/C9SE00120D (2019).
16. Xu, J. *et al.* Triple-halide wide–band gap perovskites with suppressed phase segregation for efficient tandems. *Science* **367**, 1097, doi:10.1126/science.aaz5074 (2020).
17. Tennyson, E. M., Doherty, T. A. S. & Stranks, S. D. Heterogeneity at multiple length scales in halide perovskite semiconductors. *Nat. Rev. Mater.*, doi:10.1038/s41578-019-0125-0 (2019).
18. de Quilettes, D. W. *et al.* Impact of microstructure on local carrier lifetime in perovskite solar cells. *Science* **348**, 683, doi:10.1126/science.aaa5333 (2015).

19      deQuilettes, D. W. *et al.* Tracking Photoexcited Carriers in Hybrid Perovskite Semiconductors: Trap-Dominated Spatial Heterogeneity and Diffusion. *ACS Nano* **11**, 11488-11496, doi:10.1021/acsnano.7b06242 (2017).
20      El-Hajje, G. *et al.* Quantification of spatial inhomogeneity in perovskite solar cells by hyperspectral luminescence imaging. *Energy Environ. Sci.* **9**, 2286-2294, doi:10.1039/C6EE00462H (2016).
21      Stolterfoht, M. *et al.* Visualization and suppression of interfacial recombination for high-efficiency large-area pin perovskite solar cells. *Nature Energy* **3**, 847-854, doi:10.1038/s41560-018-0219-8 (2018).
22      Doherty, T. A. S. *et al.* Performance-limiting nanoscale trap clusters at grain junctions in halide perovskites. *Nature* **580**, 360-366, doi:10.1038/s41586-020-2184-1 (2020).
23      Grancini, G. *et al.* One-Year stable perovskite solar cells by 2D/3D interface engineering. *Nat. Commun.* **8**, 15684, doi:10.1038/ncomms15684 (2017).
24      Ni, Z. *et al.* Resolving spatial and energetic distributions of trap states in metal halide perovskite solar cells. *Science* **367**, 1352-1358 (2020).
25      Jariwala, S. *et al.* Local Crystal Misorientation Influences Non-radiative Recombination in Halide Perovskites. *Joule* **3**, 3048-3060, doi:10.1016/j.joule.2019.09.001 (2019).
26      Correa-Baena, J.-P. *et al.* Homogenized halides and alkali cation segregation in alloyed organic-inorganic perovskites. *Science* **363**, 627, doi:10.1126/science.aah5065 (2019).
27      Hoke, E. T. *et al.* Reversible photo-induced trap formation in mixed-halide hybrid perovskites for photovoltaics. *Chemical Science* **6**, 613-617, doi:10.1039/C4SC03141E (2015).
28      Slotcavage, D. J., Karunadasa, H. I. & McGehee, M. D. Light-Induced Phase Segregation in Halide-Perovskite Absorbers. *ACS Energy Lett.* **1**, 1199-1205, doi:10.1021/acsenergylett.6b00495 (2016).
29      Brennan, M. C., Draguta, S., Kamat, P. V. & Kuno, M. Light-Induced Anion Phase Segregation in Mixed Halide Perovskites. *ACS Energy Lett.* **3**, 204-213, doi:10.1021/acsenergylett.7b01151 (2018).
30      Leijtens, T., Bush, K. A., Prasanna, R. & McGehee, M. D. Opportunities and challenges for tandem solar cells using metal halide perovskite semiconductors. *Nature Energy* **3**, 828-838, doi:10.1038/s41560-018-0190-4 (2018).
31      Jones, T. W. *et al.* Lattice strain causes non-radiative losses in halide perovskites. *Energy Environ. Sci.* **12**, 596-606, doi:10.1039/C8EE02751J (2019).
32      Kim, G. *et al.* Impact of strain relaxation on performance of α-formamidinium lead iodide perovskite solar cells. *Science* **370**, 108, doi:10.1126/science.abc4417 (2020).
33      Mahesh, S. *et al.* Revealing the origin of voltage loss in mixed-halide perovskite solar cells. *Energy Environ. Sci.* **13**, 258-267, doi:10.1039/C9EE02162K (2020).
34      Feldmann, S. *et al.* Photodoping through local charge carrier accumulation in alloyed hybrid perovskites for highly efficient luminescence. *Nat. Photonics* **14**, 123-128, doi:10.1038/s41566-019-0546-8 (2020).
35      Miller, O. D., Yablonovitch, E. & Kurtz, S. R. Strong Internal and External Luminescence as Solar Cells Approach the Shockley–Queisser Limit. *IEEE Journal of Photovoltaics* **2**, 303-311, doi:10.1109/JPHOTOV.2012.2198434 (2012).
36      Galisteo-López, J. F., Anaya, M., Calvo, M. E. & Míguez, H. Environmental Effects on the Photophysics of Organic–Inorganic Halide Perovskites. *J. Phys. Chem. Lett.* **6**, 2200-2205, doi:10.1021/acs.jpclett.5b00785 (2015).

37  Andaji-Garmaroudi, Z., Anaya, M., Pearson, A. J. & Stranks, S. D. Photobrightening in Lead Halide Perovskites: Observations, Mechanisms, and Future Potential. *Adv. Energy Mat.* **10**, 1903109, doi:10.1002/aenm.201903109 (2020).

38  Katahara, J. K. & Hillhouse, H. W. Quasi-Fermi level splitting and sub-bandgap absorptivity from semiconductor photoluminescence. *J. Appl. Phys.* **116**, 173504, doi:10.1063/1.4898346 (2014).

39  Stolterfoht, M. *et al.* The impact of energy alignment and interfacial recombination on the internal and external open-circuit voltage of perovskite solar cells. *Energy Environ. Sci.* **12**, 2778-2788, doi:10.1039/C9EE02020A (2019).

40  Wang, J. *et al.* Reducing Surface Recombination Velocities at the Electrical Contacts Will Improve Perovskite Photovoltaics. *ACS Energy Lett.* **4**, 222-227, doi:10.1021/acsenergylett.8b02058 (2019).

41  Mann, S. A. *et al.* Quantifying losses and thermodynamic limits in nanophotonic solar cells. *Nat. Nanotechnol.* **11**, 1071, doi:10.1038/nnano.2016.162 (2016).

42  de Mello, J. C., Wittmann, H. F. & Friend, R. H. An improved experimental determination of external photoluminescence quantum efficiency. *Adv. Mater.* **9**, 230-232, doi:10.1002/adma.19970090308 (1997).

43  Richter, J. M. *et al.* Enhancing photoluminescence yields in lead halide perovskites by photon recycling and light out-coupling. *Nat. Commun.* **7**, 13941, doi:10.1038/ncomms13941 (2016).

44  Ross, R. T. Some Thermodynamics of Photochemical Systems. *The Journal of Chemical Physics* **46**, 4590-4593, doi:10.1063/1.1840606 (1967).

45  Ledinsky, M. *et al.* Temperature Dependence of the Urbach Energy in Lead Iodide Perovskites. *J. Phys. Chem. Lett.* **10**, 1368-1373, doi:10.1021/acs.jpclett.9b00138 (2019).

46  Urbach, F. The Long-Wavelength Edge of Photographic Sensitivity and of the Electronic Absorption of Solids. *Phys. Rev.* **92**, 1324-1324, doi:10.1103/PhysRev.92.1324 (1953).

47  Piccardo, M. *et al.* Localization landscape theory of disorder in semiconductors. II. Urbach tails of disordered quantum well layers. *Phys. Rev. B* **95**, 144205, doi:10.1103/PhysRevB.95.144205 (2017).

48  Rolston, N. *et al.* Engineering Stress in Perovskite Solar Cells to Improve Stability. *Adv. Energy Mat.* **8**, 1802139, doi:10.1002/aenm.201802139 (2018).

49  Martín-Sánchez, J. *et al.* Strain-tuning of the optical properties of semiconductor nanomaterials by integration onto piezoelectric actuators. *Semiconductor Science and Technology* **33**, 013001, doi:10.1088/1361-6641/aa9b53 (2017).

50  Bioud, Y. A. *et al.* Uprooting defects to enable high-performance III–V optoelectronic devices on silicon. *Nat. Commun.* **10**, 4322, doi:10.1038/s41467-019-12353-9 (2019).

51  Hubbard, S. M. *et al.* Effect of strain compensation on quantum dot enhanced GaAs solar cells. *App. Phys. Lett.* **92**, 123512, doi:10.1063/1.2903699 (2008).

52  Man, M. K. L. *et al.* Imaging the motion of electrons across semiconductor heterojunctions. *Nat. Nanotechnol.* **12**, 36-40, doi:10.1038/nnano.2016.183 (2017).

53  Deng, S., Blach, D. D., Jin, L. & Huang, L. Imaging Carrier Dynamics and Transport in Hybrid Perovskites with Transient Absorption Microscopy. *Adv. Energy Mat.* **10**, 1903781, doi:10.1002/aenm.201903781 (2020).

54  Herz, L. M. Charge-Carrier Dynamics in Organic-Inorganic Metal Halide Perovskites. *Annual Review of Physical Chemistry* **67**, 65-89, doi:10.1146/annurev-physchem-040215-112222 (2016).

55  Yuan, M. *et al.* Perovskite energy funnels for efficient light-emitting diodes. *Nat. Nanotechnol.* **11**, 872, doi:10.1038/nnano.2016.110 (2016).

56    Lasher, G. & Stern, F. Spontaneous and Stimulated Recombination Radiation in Semiconductors. *Phys. Rev.* **133**, A553-A563, doi:10.1103/PhysRev.133.A553 (1964).
57    Wurfel, P. The chemical potential of radiation. *Journal of Physics C: Solid State Physics* **15**, 3967-3985, doi:10.1088/0022-3719/15/18/012 (1982).
58    Brüggemann, R., Schulze, P., Neumann, O., Witte, W. & Bauer, G. H. Relation between luminescence and open-circuit voltage in Cu(In,Ga)Se2 solar cells. *Thin Solid Films* **535**, 283-286, doi:10.1016/j.tsf.2012.11.039 (2013).
59    Sträter, H. *et al.* Detailed photoluminescence studies of thin film Cu2S for determination of quasi-Fermi level splitting and defect levels. *J. Appl. Phys.* **114**, 233506, doi:10.1063/1.4850955 (2013).
60    Babbe, F., Choubrac, L. & Siebentritt, S. Quasi Fermi level splitting of Cu-rich and Cu-poor Cu(In,Ga)Se2 absorber layers. *App. Phys. Lett.* **109**, 082105, doi:10.1063/1.4961530 (2016).
61    Braly, I. L., Stoddard, R. J., Rajagopal, A., Jen, A. K. Y. & Hillhouse, H. W. Photoluminescence and Photoconductivity to Assess Maximum Open-Circuit Voltage and Carrier Transport in Hybrid Perovskites and Other Photovoltaic Materials. *J. Phys. Chem. Lett.* **9**, 3779-3792, doi:10.1021/acs.jpclett.8b01152 (2018).
62    Halperin, B. I. & Lax, M. Impurity-Band Tails in the High-Density Limit. I. Minimum Counting Methods. *Phys. Rev.* **148**, 722-740, doi:10.1103/PhysRev.148.722 (1966).
63    Braly, I. L. *et al.* Hybrid perovskite films approaching the radiative limit with over 90% photoluminescence quantum efficiency. *Nat. Photonics* **12**, 355-361, doi:10.1038/s41566-018-0154-z (2018).
64    Gomez-Gonzalez, M. A. *et al.* Spatially Resolved Dissolution and Speciation Changes of ZnO Nanorods during Short-Term in Situ Incubation in a Simulated Wastewater Environment. *ACS Nano* **13**, 11049-11061, doi:10.1021/acsnano.9b02866 (2019).
65    de la Peña, F. *et al.* Hyperspy 1.5.2. (2019).
66    Kodur, M. *et al.* X-Ray Microscopy of Halide Perovskites: Techniques, Applications, and Prospects. *Adv. Energy Mat.* **10**, 1903170, doi:10.1002/aenm.201903170 (2020).
67    Virtanen, P. *et al.* SciPy 1.0: fundamental algorithms for scientific computing in Python. *Nature Methods* **17**, 261-272, doi:10.1038/s41592-019-0686-2 (2020).
68    Schnedermann, C. *et al.* Sub-10 fs Time-Resolved Vibronic Optical Microscopy. *J. Phys. Chem. Lett.* **7**, 4854-4859, doi:10.1021/acs.jpclett.6b02387 (2016).